\documentclass[a4paper]{jpconf}
\usepackage{graphicx}
\begin{document}
\title{Status of Ho\v{r}ava gravity: A personal perspective}
\author{Matt Visser}
\address{School of Mathematics, Statistics, and Operations Research \\
Victoria University of Wellington \\
Wellington, New Zealand}
\ead{matt.visser@msor.vuw.ac.nz}
\def\d{{\mathrm{d}}}
\newcommand{\scri}{\mathscr{I}}
\newcommand{\sun}{\ensuremath{\odot}}
\def\J{{\mathscr{J}}}
\def\sech{{\mathrm{sech}}}
\def\T{{\mathcal{T}}}
\def\V{{\mathcal{V}}}
\def\SS{{\mathcal{S}}}
\def\LL{{\mathcal{L}}}
\def\tr{{\mathrm{tr}}}
\def\Horava{Ho\v{r}ava}
\begin{abstract}
Ho\v{r}ava gravity is a relatively recent (Jan 2009) idea in theoretical physics for trying to develop a quantum field theory of gravity. 
It is not a string theory, nor loop quantum gravity, but is instead a traditional quantum field theory that breaks Lorentz invariance at ultra-high (presumably trans-Planckian) energies, while retaining approximate Lorentz invariance at low and medium (sub-Planckian) energies. 
The challenge is to keep the Lorentz symmetry breaking controlled and small --- small enough to be compatible with experiment. 
I will give a very general overview of what is going on in this field, paying particular attention to the disturbing role of the scalar graviton. 

\end{abstract}

\section{Introduction} 

Is Lorentz symmetry truly fundamental? Or is it just an ``accidental" low-momentum emergent symmetry? Opinions on this issue have undergone a radical mutation over the last few years. Historically, (for good experimental and theoretical reasons), Lorentz symmetry was considered absolutely fundamental --- not to be trifled with --- but for a number of independent reasons the modern viewpoint is more nuanced. 
\begin{itemize}
\item 
First, there is now a well developed phenomenological framework that permits precision tests of Lorentz invariance. (See for instance~\cite{Mattingly:05, Mattingly:08, Jacobson:01, Jacobson:05}.) While all observational data to date are compatible with exact Lorentz invariance, the existence of this theoretical framework shows that at least some (limited) types of Lorentz symmetry breaking are compatible with current observations. 
\item
Second, there are very many and varied theoretical scenarios that suggest (but do not unavoidably guarantee) that at sufficiently high energies (typically  trans-Planckian energies) Lorentz symmetry might eventually break down. (See for instance~\cite{Kostelecky1, Kostelecky2}.)
\end{itemize}
Faced with this theoretical situation it makes very good sense to ask:
 \begin{itemize}
\item 
What if anything are the benefits of Lorentz symmetry breaking? 
 
\item
What can we do with it? 
 
\item
Why should we care?
 
\item
Where are the bodies buried?
 \end{itemize}
In this regard, one of the more significant developments of the last decade was the appearance in early 2009 of a paper by \Horava~\cite{Horava1} which laid out a strategy for making a virtue of Lorentz symmetry breaking by \emph{using} it as a key ingredient in developing a superficially plausible quantum field theoretic candidate for a theory of quantum gravity.  \Horava's model is not a string theory, nor loop quantum gravity, but is instead a traditional (point-particle) quantum field theory that breaks Lorentz invariance at ultra-high (presumably trans-Planckian) energies. Since the mere existence of any such field-theoretic model (independent of its actual phenomenological viability) flew in the face of much of the ``perceived wisdom'' regarding the features one might reasonably expect of any realistic theory of ``quantum gravity'', there was very rapid and extremely extensive interest in~\Horava's proposal. 

As of March 2011, \Horava's original paper~\cite{Horava1} has been cited over 425 times, with some 65\% of the citing articles being published. If we add the closely related articles~\cite{Horava0, Horava2} this rises to some 450 articles, with again some 65\% of the citing articles being published. This is indicative of a very highly active research field. On the other hand, there is at least one note of caution that should be raised: The unpublished 35\% of citing articles cannot all be attributed simply to time-lag, nor to non-journal conference-proceedings articles.  The two main reasons underlying this state of affairs seem to be:
\begin{itemize}
\item 
There has been somewhat of a tendency to charge full steam ahead with applications, (typically cosmological), without first fully understanding the foundations of the model.
\item
For that matter, there is still considerable disagreement as to what precise version of \Horava's model is ``best''. 
\end{itemize}
Consequently, when reading the relevant literature, a certain amount of caution is advisable. In the current article I will give a rather personal interpretation of the current state of affairs in \Horava{} gravity, with some indications of possible future directions.

\section{Basic framework: Field theory without gravity} 

My interpretation of the central idea of \Horava{} gravity is this: One abandons ultra-high-energy Lorentz invariance as fundamental, recognizing that phenomenologically  one need ``merely'' attempt to recover an approximate low-energy Lorentz invariance. Typical dispersion relations are then of the form
\begin{equation}
 \omega =  \sqrt{m^2+k^2 + {k^4\over K^2} + \dots} 
\end{equation}
This point of view is nicely compatible with the ``analogue spacetime'' programme~\cite{ergo, lrr}, which is one reason I personally was very quick to get involved. More precisely, in condensed matter language, a so-called ``critical'' Lifshitz point in $(d+1)$ dimensions corresponds to a 
 dispersion relation which satisfies
\begin{equation}
\omega \to  k^d \qquad \hbox{as} \qquad k \to \infty.
\end{equation}
To recover Lorentz invariance, at ``low'' momentum (but still allowing {$k \gg m$})  the dispersion relation should satisfy
\begin{equation}
\omega \to  \sqrt{m^2+k^2} \qquad \hbox{as} \qquad {k \to 0}.
\end{equation}
Note that every quantum field theory (QFT) regulator currently known to mankind either breaks Lorentz invariance explicitly (e.g.~lattice QFT), or does something worse, often something outright unphysical. For example: Pauli--Villars violates unitarity; Lorentz-invariant higher-derivatives violate {unitarity}; 
dimensional regularization is at best a purely {formal} trick with no direct physical interpretation (and which requires a Zen-like approach to gamma matrix algebra).
The standard viewpoint is this:
If the main goal is efficient computation in a corner of parameter space that we experimentally know to be Lorentz invariant to a high level of precision, then by all means, go ahead and develop a Lorentz-invariant perturbation theory  with an unphysical regulator
--- hopefully the unphysical aspects of the computation can first be isolated, and then banished by renormalization.
This is exactly what is done, (very efficiently and very effectively), 
in the ``standard model of particle physics''. (See for instance~\cite{Weinberg}.)

\Horava's approach can be interpreted by taking on a non-standard viewpoint:
If  one has reason to suspect that Lorentz invariance might ultimately break down at ultra-high (presumably trans--Planckian) energies, 
then a different strategy suggests itself.
Maybe one could \emph{use} the Lorentz symmetry breaking as part of the QFT regularization procedure?
Could we then at least keep intermediate parts of the QFT calculation ``physical''?
(Note that ``physical'' does not necessarily mean ``realistic'', it just means we are not violating fundamental tenets of quantum physics at intermediate stages of the calculation.)

Consider for example a specific ``physical'' but Lorentz-violating regulator based on the dispersion relation
\begin{equation}
\omega^2 = m^2 + k^2 + {k^4\over K_4^2} + {k^6\over K_6^4}.
\end{equation}
We call this a ``trans--Bogoliubov'' dispersion relation~\cite{lrr}, based on analogy with the standard condensed-matter Bogoliubov dispersion relation:
\begin{equation}
\omega^2 = k^2 + {k^4\over K^2}.
\end{equation}
The corresponding QFT propagator (momentum-space Green function) is
\begin{equation}
G(\omega, k) = {1\over \omega^2 - \left[ m^2 + k^2 + {k^4/ K_4^2} + {k^6/K_6^4}\right]}.
\end{equation}
Note the rapid fall-off as spatial momentum {$k\to\infty$}.
This improves the behaviour of the integrals encountered in Feynman diagram calculations (QFT perturbation theory).
Specifically: In any (3+1) dimensional scalar QFT, with arbitrary polynomial self-interaction, this is enough (after Wick rotation and normal ordering), 
to keep all individual Feynman diagrams \emph{finite}. For details see~\cite{Visser:QFT}, the original \Horava{} article~\cite{Horava1}, and related work by Anselmi~\cite{Anselmi1, Anselmi2}. More generally, with language again borrowed from condensed matter,
for a Lifshitz point of order $z$ in  $(d+1)$ dimensions we have the dispersion relation:
\begin{equation}
\omega^2 = m^2 + k^2 + { \sum_{n=2}^z g_n \; {k^{2n}\over K^{2n-2}} }.
\end{equation}
The equivalent QFT propagator [Green function] is now
\begin{equation}
G(\omega, k) = {1\over \omega^2 - \left[  m^2+k^2 + { \sum_{n=2}^z g_n \; {k^{2n}/ K^{2n-2} }} \right]}.
\end{equation}
Key theoretical results are that:
\begin{itemize}
\item
In a  $(d+1)$ dimensional scalar QFT with $z=d$, and arbitrary polynomial self-interaction, this is enough (after Wick rotation and normal ordering) to keep all Feynman diagrams individually \emph{finite}~\cite{Horava1, Visser:QFT}.
\item
Gravity is a little trickier, but you can at least argue for power-counting renormalizability of the resulting QFT~\cite{Horava1, Visser:QFT2}.
\end{itemize}
These results are unexpected, seriously unexpected. And yes, there are still significant technical difficulties, (of which more anon).

\section{Basic framework: Gravity} 
Consider the standard ADM decomposition for $(d+1)$ dimensional gravity:
\begin{equation}
\SS =  {\int \sqrt{-g_{d}} \;\; N \;\;\left\{  \tr[K^2] - \tr[K]^2 + {}^{(d)} R \right\} \;\; \d^{d} x\; \d t}.
\end{equation}
Here one has split spacetime into space+time, while $ {}^{(d)} R$ is intrinsic curvature of space, and $K$ is the extrinsic curvature of space in spacetime. 
On the one hand this ADM decomposition leads to ``canonically quantized gravity'' 
(with all its problems), and on the other hand it is classically very useful for numerical relativity.
The key to \Horava{} gravity is now to develop a non-standard extension of the ADM formalism: 
\begin{itemize}
\item 
Choose a ``preferred foliation''.
\item
Decompose $\LL =  \hbox{(kinetic term)} - \hbox{(potential term)}$. 
\item 
Add extra ``kinetic'' and ``potential'' terms, beyond what you expect from Einstein--Hilbert.
\item
You cannot now reassemble $\LL$ into a simple $^{(d+1)} R$.
\item
You have now implicitly reintroduced the aether, more on this below.
\end{itemize}
Specifically:
\begin{itemize}
\item 
Consider the ``kinetic energy''
\begin{equation}
 \T(K) = {g_K}     \left\{ (K^{ij} K_{ij} - K^2)+ \xi K^2 \right\} =  {g_K}     \left\{ K^{ij} K_{ij} - \lambda K^2 \right\} ,
\end{equation}
noting that standard general relativity would enforce $\xi\to0$, (that is $\lambda\to1$). Take the kinetic action to be
\begin{equation}
\SS_K =    \int \T(K) \; \d V_{d+1} =    \int \T(K) \sqrt{g} \; N\; \d^d x \; \d t.
\end{equation}
This contains only two time derivatives (hiding in $K$) --- this is good, and is necessary (though not by itself sufficient) to suppress ghosts. 
\item 
There is also a hidden ``scalar graviton'' when $\xi\neq0$ --- this is potentially very bad, and I will have much more to say on this below. 
\item 
Now consider the most general ``potential energy'' in $(d+1)$ dimensions:
\begin{equation}
\SS_\V = 
\int   \V(g, N) \; \sqrt{g} \; N \; \d^d x \; \d t,
\end{equation}
where $\V(g, N)$ is some scalar built out of the spatial metric and the lapse function, and their spatial derivatives.
\end{itemize}

\section{Simplifying ansatz\"e} 
It is roughly at this stage that Ho\v{r}ava makes his two great simplifications:

\begin{itemize}
\item Detailed balance.
\item Projectability.
\end{itemize}
Even after almost two  years it is still somewhat unclear whether these are \emph{only} ``simplifying ansatz\"e'' or whether they are in some sense fundamental to Ho\v{r}ava's model. In particular, Thomas Sotiriou, Silke Weinfurtner, and I have argued that ``detailed balance'' is not fundamental~\cite{SVW1, SVW2}, and we have been carefully thinking about the issue of ``projectability''.

\paragraph{Detailed balance:}
What is Ho\v{r}ava's detailed balance condition? It is the assertion that the potential $\V(g,N)$ is a perfect square.
That is, there is a ``pre-potential'' $W(g,N)$ such that:
\begin{equation}
{\V(g,N) = \left( g^{ij} \; {\delta W\over\delta g_{jk} } \; g^{kl} \;  {\delta W\over\delta g_{li} } \right) .} 
\end{equation}
This simplifies some features of Ho\v{r}ava's model, it makes other features much worse.
In particular, if you assume Ho\v{r}ava's detailed balance, and try to recover the Einstein--Hilbert action in the low-energy regime,  then:
\begin{itemize}
\item You are forced to accept a non-zero cosmological constant of the wrong sign.
\item You are forced to accept intrinsic parity violation in the purely gravitational sector.
\end{itemize}
Now the second item I could live with, albeit uncomfortably, (observationally there is not the slightest hint of parity violation in gravitational physics), but dealing with the the first item will require some mutilation of detailed balance anyway, so we might as well go the whole way and discard detailed balance entirely.

\paragraph{Projectability:} What is Ho\v{r}ava's projectability condition?
\begin{equation}
{N(x,t) \to N(t)} \quad  (\hbox{whence we can subsequently set } N(t) \to 1).
\end{equation}
In standard general relativity the ``projectability condition''  can always be enforced locally as a gauge choice.
Furthermore for ``physically interesting'' solutions of general relativity it seems that this can always be done (more or less) {globally}. 
For instance:
\begin{itemize}
\item 
For the Schwarzschild spacetime this ``projectability condition'' holds globally in Painlev\'e--Gullstrand coordinates. 
\item 
For the Reissner--Nordstr\"om  spacetime this ``projectability condition'' holds for $r\geq Q^2/2m$  in Painlev\'e--Gullstrand coordinates,
 (which covers the physically interesting region of the spacetime down to some point deep inside the inner horizon).
\item 
For the Kerr spacetime this condition holds globally (for the physically interesting  chronology respecting $r>0$ region) in Doran coordinates. (So that the geometry is ``projectable'' down  to some point deep inside the inner horizon).
\item  
The FLRW cosmologies also automatically satisfy this ``projectability condition''. 
\end{itemize}
For this purely pragmatic reason we decided to put ``projectability'' off to one side for a while, and first deal with ``detailed balance''~\cite{SVW1, SVW2} (see~\cite{WSV:nutshell} and~\cite{Sotiriou:status} for more recent discussions).

\paragraph{Pragmatic projectable parity preserving model:} Abandoning detailed balance, but retaining projectability and parity invariance, 
it follows that $\V(g)$ must be built out of scalar curvature invariants --- and so it is calculable in terms of the Riemann tensor and its derivatives.
This tells us the marginal operators in the potential energy must be constructible from objects of the form
\begin{equation}
{ \left\{  (\hbox{Riemann})^d,  [(\nabla\hbox{Riemann})]^2 \hbox{(Riemann})^{d-3} , \hbox{etc...}   \right\} }.
\end{equation}
In $d+1$ dimensions this is a long but finite list. 
Note that all of these theories should be well-behaved as QFTs (at least in terms of being power-counting renormalizable).
All of these theories should have (in condensed matter language) ``$z=d$ Lifshitz points''.
In the specific case $d=3$ we have the short and rather specific list:
\begin{equation}
\Big\{  (\hbox{Riemann})^3,  [\nabla(\hbox{Riemann})]^2,    
 (\hbox{Riemann})  \nabla^2(\hbox{Riemann}),   \nabla^4(\hbox{Riemann}) \Big\}.
\end{equation}
But in 3 dimensions the Weyl tensor automatically vanishes,
so we can always decompose the Riemann tensor into the Ricci tensor, Ricci scalar, plus the metric. 
Thus we need only consider the much simplified list:
\begin{equation}
\Big\{  (\hbox{Ricci})^3,  [\nabla(\hbox{Ricci})]^2,    
 (\hbox{Ricci})  \nabla^2(\hbox{Ricci}),   \nabla^4(\hbox{Ricci}) \Big\}. \quad
\end{equation}
Once you look at all the different ways the indices can be wired up this is still relatively messy.
In (3+1) dimensions there are only five independent marginal terms (renormalizable by power counting):
\begin{equation}
 R^3, \quad R R^{i}{}_{j} R^{j}{}_i, \quad  R^i{}_j R^j{}_k R^k{}_i; \quad 
R \; \nabla^2 R, \quad  \nabla_i R_{jk} \, \nabla^i R^{jk}.
\end{equation}
Now add all possible lower-dimension terms (relevant operators, super-renormalizable by power-counting):
\begin{equation}
 {1};
\qquad
\; \;   {R};
\qquad
\; \;   {R^2}; \; \; {R^{ij} R_{ij}}.
\end{equation}
If one absolutely insists on destroying parity invariance ``by hand'' one could add the relevant (super-renormalizable) operator:
\begin{equation}
R^{ij} C_{ij},
\end{equation}
where $C_{ij}$ is the Cotton tensor (which is parity odd since it depends on the 3-dimensional Levi--Civita tensor for its definition).
Suppressing  parity violation, this results in a potential $\V(g)$ with nine terms and nine independent coupling constants. 
The Einstein--Hilbert piece of the action is
\begin{equation}
\SS_\mathrm{EH} =   \zeta^2 \int   \left\{  (K^{ij} K_{ij} - K^2) +  R  -  g_0\, \zeta^2 \right\} \sqrt{g} \; N\; \d^3 x \; \d t.
\end{equation}
The ``extra'' Lorentz-violating terms become:
\begin{eqnarray}
\SS_\mathrm{LV} &=&  \zeta^2 \int   \Big\{ \xi\, K^2 -     g_2 \,\zeta^{-2}\,R^2 -  g_3 \, \zeta^{-2}\, R_{ij} R^{ij} 
- g_4 \,  \zeta^{-4}\,R^3 - g_5 \,\zeta^{-4}\, R (R_{ij} R^{ij})
\nonumber\\
&&
\qquad
- g_6 \,\zeta^{-4}\, R^i{}_j R^j{}_k R^k{}_i 
- g_7\,\zeta^{-4}\, R \nabla^2 R
 - g_8 \,\zeta^{-4}\, \nabla_i R_{jk} \, \nabla^i R^{jk}
\Big\} \sqrt{g} \; N\; \d^d x \; \d t. \qquad
\end{eqnarray}
From the normalization of the Einstein--Hilbert term:
\begin{equation}
(16 \pi G_\mathrm{Newton})^{-1} = \zeta^2;  \qquad\qquad \Lambda  =  {g_0 \, \zeta^2\over2};
\end{equation}
so that $\zeta$ is identified as the Planck scale.  
The cosmological constant is determined by the free parameter $g_0$, and observationally $g_0 \sim 10^{-123}$,
(renormalized after including  vacuum energy contributions).  In particular, the way we have set this up we are free to choose the Newton constant and cosmological constant independently (and so to be compatible with observation). 
Key features of the model are:
\begin{itemize}
\item 
The Lorentz violating term in the kinetic energy leads to an extra scalar mode for the graviton, with fractional $O(\xi)=O(\lambda-1)$ effects at all momenta.  
Phenomenologically, this behaviour is potentially dangerous and should be carefully investigated. 
\item
The  various Lorentz-violating terms in the potential become comparable to the spatial curvature term in the Einstein--Hilbert action for physical momenta of order
\begin{equation}
\zeta_{\{2,3\}} = {\zeta\over\sqrt{ |g_{\{2,3\}}| }};  \qquad 
\zeta_{\{4,5,6,7,8\}} = {\zeta\over\sqrt[4]{ |g_{\{4,5,6,7,8\}}| }}.
\end{equation}
\item 
The Planck scale $\zeta$ is divorced from the various Lorentz-breaking scales $\zeta_{\{2,3,4,5,6,7,8\}}$.
\item
One can drive the Lorentz breaking scale arbitrarily high by suitable adjustment of the dimensionless couplings $ g_{\{2,3\}}$ and  $g_{\{4,5,6,7,8\}}$.
\end{itemize}
Based on his intuition coming from ``analogue spacetimes'', Grisha Volovik has for many years  been asserting that the Lorentz-breaking scale should be much higher than the Planck scale~\cite{Volovik:2007, Volovik:2008, Volovik:2009, Volovik:2010}.  This model very naturally implements that idea in a concrete and explicit manner.

\section{Potential problems} 
Where are the bodies buried? 
\begin{itemize}
\item 
\emph{Projectability:} Among other things, this yields a spatially integrated Hamiltonian constraint rather than a super-Hamiltonian constraint. This modifies the equations of motion away from those of standard general relativity. Some authors have made a virtue out of necessity by using the Hamiltonian/super-Hamiltonian distinction to implement ``dark dust'' at the level of the equations of motion~\cite{dark-dust}. 
\item 
\emph{Prior structure:} Is the preferred foliation ``prior structure''?  Or is it dynamical? Can it be made dynamical by giving up projectability? Can  it be viewed as a variant of Einstein--aether theory? (See~\cite{aether1, aether2, aether3, aether4}.) Note that I particularly wish to avoid using the phrase ``diffeomorphism invariance'' since many authors fail to distinguish between ``active'' and ``passive'' diffeomorphisms. Invariance under active diffeomorphisms is equivalent to the absence of prior structure, and \Horava's model (because of the preferred foliation) is not diffeomorphism invariant in this active sense \emph{as long as the preferred foliation is non-dynamical}. On the other hand, invariance under passive diffeomorphisms is just coordinate re-parameterization invariance, and \emph{any} theory can be rendered coordinate re-parameterization invariant by introducing enough prior structure. In particular, \Horava's models are  \emph{always} diffeomorphism invariant in this passive sense. 
\item 
\emph{Scalar graviton:} As long as $\xi\neq0$ there is a spin-0 scalar graviton, 
 in addition to the spin-2 tensor graviton. This is disturbing for a number of reasons, both theoretical and phenomenological.
\item
\emph{Hierarchy problem}? 
Even if the theory is finite, it may still require fine tuning  in order to be compatible with observation.  Finite does not necessarily mean realistic.  It is worth repeating the well-known fact that even finite QFTs still need to be renormalized: One still needs to go through the process of rewriting the bare parameters in terms of the renormalized parameters that are actually available for experimental observation. 
\item
\emph{Beta functions}? Can one go beyond power-counting renormalizability? (See for instance~\cite{Orlando1, Orlando2}, though note that those authors used detailed balance as an essential part of their formalism.)What is the renormalization group (RG) flow?
Explicit calculations have so far been extremely limited. (See for instance~\cite{Giribet}.)
\end{itemize}
All of these issues have the potential for raising violent conflicts with empirical reality.
\begin{table}[htb]
\begin{center}
\begin{tabular}{cc}
\hline
 \hbox to 5 cm{\hfil Conformal point\hfil} & \hbox to 5 cm{\hfil General relativity \hfil}  \\
\hline
  $\lambda= 1/3$ & $\lambda=1$  \\ 
 \hline
 \vphantom{\Bigg{|}}
 \dotfill $\star\longrightarrow\longrightarrow\longrightarrow\longrightarrow\!\!\!$&$\!\!\!\longrightarrow\longrightarrow\longrightarrow\longrightarrow\star$\dotfill  \\
 \hline
  $\xi = 2/3$ &  $\xi = 0$ \\
 \hline
\end{tabular}
\end{center}
\caption{Conjectured RG flow in \Horava's original model: From conformal invariance in the UV to general relativity in the IR. Note that the RG flow is confined to a region where the scalar mode is a ghost (negative kinetic energy). }
\end{table}

\noindent
In particular note that:
\begin{itemize}
\item 
In the projectable model the unwanted scalar mode has negative kinetic energy all the way from the conformal point ($\xi = 2/3$, $\lambda=1/3$) to the GR point ($\xi=0$, $\lambda=1$). The scalar mode is elliptic (unstable) in this entire region. The often conjectured RG flow from the conformal point $\lambda=1/3$ in the UV to the GR point $\lambda =1$ in the IR is not viable.
\item 
From a theoretical perspective the scalar mode also leads to undesirable behaviour such as exhibiting the potential for strong coupling at low energies~\cite{Charmousis, Kimpton, Padilla}, instabilities, and over-constrained evolution. (See for instance~\cite{WSV:nutshell, Sotiriou:status} for overviews).

\item
In addition, the spin-0 scalar graviton is potentially dangerous for a number of purely phenomenological reasons:
\begin{itemize}
\item Binary pulsar?  (The scalar mode is expected to provide an extra energy-loss mechanism. But how significant is this energy loss?)
\item PPN physics? (In principle the scalar mode might affect the PPN parameters and modify solar system physics. Could it be detected via precise solar system tests?)
\item E\"otv\"os experiments? (Scalar modes generically affect the universality of free-fall. Can this effect in \Horava-like models be suitably suppressed to remain compatible with observations?)
\end{itemize}
I emphasize that if one wishes to attack any of these phenomenological issues one will need to make a very precise commitment as to which particular version of \Horava's model one is dealing with, and also carefully check it for unrelated problems. (See for instance~\cite{Kimpton} for a recent analysis, though that article is more focussed on the non-projectable version of \Horava's  model.)
\end{itemize}
Considerable careful thought on these issues is still needed.

\section{Potential solutions} 

Several options for improving the behaviour of \Horava-like models have been explored. 

\paragraph{Healthy extensions:} One way of dealing with at least some of the issues arising from the scalar mode is through the so-called  ``healthy extension''  (also called the ``consistent extension'') of \Horava's model~\cite{Blas0,Blas1,Blas2}. The central idea of this ``consistent extension'' is to make the ``preferred foliation'' dynamical. Effectively one is then dealing with a stably causal spacetime with a preferred dynamical ``cosmic time''.
\begin{itemize}
\item 
In the ``healthy extension'' models one enforces $\lambda>1$ ($\xi<0$) to avoid negative kinetic energies for the scalar mode.
\item 
One then needs to go beyond projectability to make the spin zero mode hyperbolic, thus greatly increasing the number of  terms appearing in the potential energy part of the action, and greatly increasing the overall complexity of the model. (Thus one is considering a version of the general \Horava{} model, with neither detailed balance nor projectability.) 
\item 
To obtain desirable low-energy limit one has to \emph{postulate} that the parameter $\xi$ has RG flow down to $\xi=0$ from below. (That is $\xi \to 0^-$; equivalently $\lambda\to 1^+$.) 
\item
Because the preferred foliation is now dynamical, the healthy extension models are related to so-called Einstein-aether models. (See in particular~\cite{aether4}, and~\cite{aether1, aether2, aether3} for more background.)
\item
Somewhat unusually, when linearized around flat space the scalar mode (though not the spin-2 graviton) acquires a rational polynomial dispersion relation
\begin{equation}
\omega^2 = {p(k^2)\over q(k^2)},
\end{equation}
where in 3 space dimensions $\mathrm{degree}[p] = \mathrm{degree}[q] + 3$.  
\item
Recently it has been noted that this picture qualitatively continues to hold in (2+1) dimensions, though then $\mathrm{degree}[p] = \mathrm{degree}[q] + 2$. Furthermore the scalar mode is in (2+1) dimensions the \emph{only} local degree of freedom~\cite{SVW:2+1}. This makes (2+1) \Horava{} gravity a useful toy model to work with, insofar as it captures the essential physics of the scalar mode, but without the complications attendant on the presence of the spin 2 graviton. 
\item
Strong coupling may still be an issue~\cite{Sotiriou:status, Papazoglou, Blas:2009, Wang:2010}.

\end{itemize}
\begin{table}[thb]
\begin{center}
\begin{tabular}{cc}
\hline
 \hbox to 5 cm{\hfil Conformal point\hfil} & \hbox to 5 cm{\hfil General relativity \hfil}  \\
\hline
  $\lambda= 1/3$ & $\lambda=1$  \\ 
 \hline
 \vphantom{\Bigg{|}}
 \dotfill $\star${\dotfill}$\!\!\!\!\!$&$\!\!\!\!\!$ {\dotfill}$\star\longleftarrow\longleftarrow\longleftarrow\longleftarrow $\\
 \hline
  $\xi = 2/3$ &  $\xi = 0$ \\
 \hline
\end{tabular}\end{center}
\caption{Conjectured RG flow in ``healthy extension''/``consistent extension" models.  The RG flow is now postulated to come from a region where both the scalar mode and spin 2 graviton have positive kinetic energy. }
\end{table}

\paragraph{Analogue spacetimes:} Another option is to search for modifications of \Horava's original models based on hints coming from the ``analogue spacetime'' programme. A model based on a Bose liquid defined on a lattice has been proposed by Xu and \Horava~\cite{Horava:Bose}.  Unfortunately, while this particular model has interesting stability properties, it typically violates Lorentz invariance at low energy, which makes realistic model building implausible. On the other hand, the general idea of being guided by the analogue spacetime programme may prove useful in slightly different contexts.

\paragraph{Extra symmetries:} One can try to tame the scalar mode by adding an extra gauge symmetry. See for instance~\cite{Horava:extra}. 
In that particular model the extra symmetry does \emph{not} seem to be related to diffeomorphism invariance in any way, and its physical significance is thereby somewhat obscured. 

\paragraph{Other:} Other more radical proposals are to somehow merge \Horava-like models with causal dynamical triangulations (CDTs) in the UV. See particularly~\cite{CDT-meets-Horava}, with additional background discussion in~\cite{Horava2, Horava:extra}.

\paragraph{Summary:} Overall, the choices seem to be:
\begin{itemize}
\item Approximately decouple the scalar mode?
\item Kill the scalar mode with more symmetry?
\item Kill the scalar mode with less symmetry? Impose even more drastic a priori restrictions on the metric?
(Even more restrictions than projectability.)
\end{itemize}
For various reasons, none of these approaches is as yet fully satisfying.

\section{Discussion}

The (generalized) \Horava-like models naturally provide a class of tempting models, of varying degrees of complexity and physical viability, that are rather interesting to work on. My personal view is that considerably more work needs to go into understanding the fundamentals of these models, specifically the role played by the scalar mode.  Without a deeper understanding of the fundamental framework one is operating in,  detailed phenomenological studies (and in particular specific applications to cosmology and astrophysics) are simply premature. Specifically, one needs more than hand-waving ``of course it runs to general relativity in the IR'' arguments. There may be subtle (or even not so subtle) qualitative deviations from general relativity due to the preferred foliation, and really pinning that issue down would be a good idea before investing more time on detailed applications.  Overall, this is still a very active field. While the initial feeding frenzy has somewhat subsided, considerable ongoing interest remains.  There are also some very real physics challenges remaining. 

\ack

This research was supported by the Marsden Fund administered by the Royal Society of New Zealand. I also wish to thank the CSIC-IAA (Granada) for hospitality, and to thank Thomas Sotiriou and Silke Weinfurtner for their comments.

\section*{References}

\end{document}